\documentclass[aps,preprint,preprintnumbers,amssymb,amsfonts,groupedaddress,superscriptaddress]{revtex4}

\usepackage{graphicx}
\usepackage{subfigure}
\usepackage{bm}% bold math
\usepackage{color}

\newcommand{\bmath}{\begin{mathletters}}
\newcommand{\emath}{\end{mathletters}}
\newcommand{\be}{\begin{eqnarray}}
\newcommand{\ee}{\end{eqnarray}}
\newcommand{\ba}{\begin{array}}
\newcommand{\ea}{\end{array}}

\newcommand{\de}{\delta}

\newcommand{\Ga}{\Gamma}

\newcommand{\h}{\hbar}
\newcommand{\pr}{\prime}

\begin{document}
\title{Generic Mechanism of Optimal Energy Transfer Efficiency:
A Scaling Theory of the Mean First Passage Time in Exciton Systems
}\textbf{}
\author{Jianlan Wu}
\affiliation{Department of Chemistry, MIT, 77 Massachusetts Ave, Cambridge, MA, 02139, USA}
\affiliation{Physics Department, Zhejiang University, 38 ZheDa Road, Hangzhou, Zhejiang, 310027, China}
\author{Robert J. Silbey\footnote{Dedicated to the memory of Prof. Robert J. Silbey}}
\affiliation{Department of Chemistry, MIT, 77 Massachusetts Ave, Cambridge, MA, 02139, USA}
\author{Jianshu Cao\footnote{Electronic address: jianshu@mit.edu}}
%\email{jianshu@mit.edu}
\affiliation{Department of Chemistry, MIT, 77 Massachusetts Ave, Cambridge, MA, 02139, USA}

\date{\today}

\begin{abstract}
An asymptotic scaling theory is presented using the conceptual basis
of trapping-free subspace (i.e., orthogonal subspace) to establish
the generic mechanism of optimal efficiency of excitation energy
transfer (EET) in light-harvesting systems. Analogous to Kramers'
turnover in classical rate theory, the enhanced efficiency in the
weak damping limit and the suppressed efficiency in the strong
damping limit define two asymptotic scaling regimes, which are
interpolated to predict the functional form of optimal efficiency of
the trapping-free subspace. In the presence of static disorder, the
scaling law of transfer time with respect to dephasing rate changes
from linear to square root, suggesting a weaker dependence on the
environment. Though formulated in the context of EET, the analysis
and conclusions apply in general to open quantum processes,
including electron transfer, fluorescence emission, and heat
conduction.

\end{abstract}

\maketitle

The optimization of the excitation energy transfer (EET) process
presents a challenge for understanding photosynthetic systems as
well as for designing efficient solar energy devices.
Multi-dimensional spectra have allowed detailed probes of EET
dynamics, revealing signatures of quantum
coherence~\cite{engel2007:Nature,lee2007:Science,collini2010:Nature,engel}.
In contrast to the common belief that noise retards motion and
coherence enhances mobility, the EET efficiency can be optimized at
an intermediate level of noise, leading to the notion of
noise-enhanced
energy transfer~\cite{gaab2004:JChemPhys,vlaming2007:JCP,robentrost2009:NewJPhys,caruso2009:JChemPhys,cao2009:JPCA,wu2010:NewJPhys}. %olaya-castro2008:PhysRevB,
Fermi's golden rule rate provides a simple interpretation and
suggests that stochastic resonance between the donor
and acceptor enhances EET efficiency~\cite{caruso2009:JChemPhys,cao2009:JPCA,wu2010:NewJPhys}. %,forster1948,forster}.
Another possible mechanism is that noise suppresses destructive
interference between
pathways~\cite{caruso2009:JChemPhys,cao2009:JPCA}. The situation is
further complicated by the findings that initial preparation,
coherence of incident photons, site energy, spatial correlation,
static disorder, and various approximations invoked in quantum
master
equations  %~\cite{haken1972, haken1973:ZPhys,grover1971:JChemPhys,tanimura1989:JPSJ,cao1997,mukamel,ishizaki2009:PNAS}
can all play a role in establishing optimal
 efficiency~\cite{wu2010:NewJPhys,Whaley:NatPhys,Moix2011,Mohseni2011:arXiv2,Brumer2011:IncidentPhoton,Mancal2010}.
Therefore, a general mechanism for optimization in an arbitrary EET
system accompanying all these effects is clearly needed but has not
yet been formulated.

In this Letter, we utilize the concept of trapping-free subspace (i.e., orthogonal subspace)
to bring together all of the above considerations into a unified framework, that
allows us to establish asymptotic scaling relations under both
dynamic and static disorder and identify the generic behavior of
optimal EET efficiency. The definition of a trapping-free subspace is a
generalization of the invariant subspace~\cite{caruso2009:JChemPhys}
and is closely related to the
concept of decoherence-free subspace in quantum
information~\cite{whaley1998:PRL}.

\begin{figure}[htp]
 \includegraphics[width=0.9\columnwidth]{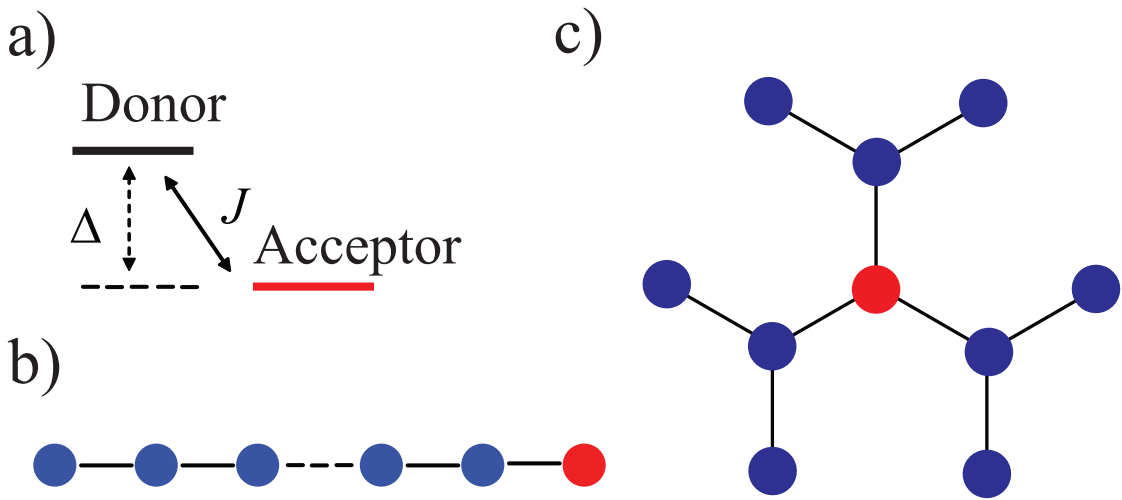}
  \caption{(color online) Various EET systems with the trapping free subspace:
  a) A donor-acceptor system with an energy mismatch $\Delta$ larger than the site-site coupling $J$ ($\Delta\gg J$).
  b) A homogeneous $N$-site ($N\gg 1$) chain where the site-site coupling decays exponentially with distance.
  %c) A top view of a perfect ring with the trap site in the center, a qualitative description of the LH1 system~\cite{schulten}.
  c) A schematic diagram of a two-generation three-fold
  dendrimer~\cite{supritz2006:JLumin}.} %{jiang1997:Nature,supritz2006:JLumin}. }
  \label{fig00}
\end{figure}

{\it Model.} --- We consider a light-harvesting EET system
(see examples in Fig.~\ref{fig00}) described by the quantum master
equation for the reduced density matrix of the single excitation manifold~\cite{cao2009:JPCA},
%\be
%\dot{\rho}(t) &=& -\mathcal{L}\rho(t)\no \\
% &=& -\left[\mathcal L_{\mathrm{sys}}+\mathcal L_{\mathrm{decay}}
%+\mathcal L_{\mathrm{trap}}+\mathcal
%L_{\mathrm{dissip}}\right]\rho(t),
%    \ee
\be
\dot{\rho}(t) = -{\mathcal L} \rho(t).
\ee
Here, the Liouville superoperator $\mathcal L=\mathcal L_{\mathrm{sys}}+\mathcal L_{\mathrm{dissip}}
+\mathcal L_{\mathrm{trap}}+\mathcal L_{\mathrm{decay}}$ comprises four terms,
each describing a distinct dynamic process: (i) $\mathcal
L_{\mathrm{sys}}\rho(t)=(i/\h)[H_\mathrm S, \rho(t)]$, the dynamics
of the isolated system, where $H_\mathrm S$ is the system
Hamiltonian; (ii) $\mathcal L_{\mathrm{dissip}}$, the exciton re-distribution and
dephasing within the single-excitation manifold due to the interaction with the surrounding environment;
(iii) $\mathcal L_{\mathrm{trap}}$, the trapping of excitation
energy at the reaction center for the production of chemical energy;
(iv)  $\mathcal L_{\mathrm{decay}}$, the decay of the
excitation energy to ground state in the form of heat or a photon.
The Liouville superoperator $\mathcal L$ becomes invertible because of
the two irreversible energy depletion terms,
$\mathcal L_{\mathrm{trap}}$ and $\mathcal L_{\mathrm{dissp}}$.
A basic property of an EET system is its energy transfer efficiency,
\be
q = \mathrm{Tr}\int_0^\infty \mathcal L_{\mathrm{trap}} \rho(t) dt,
\label{eq02}
\ee
where $\mathrm{Tr}$ denotes the trace over states.
For an efficient EET system such as a %photosynthetic
light-harvesting protein complex, its nearly unit energy transfer efficiency, $q\sim1$,
implies a clear time-scale separation between the decay process and the trapping
process. Under this condition and
with a homogeneous decay rate $k_d$,
the transfer efficiency is given, to a good approximation~\cite{cao2009:JPCA}, by
\be
q \approx \frac{1}{1+k_d \langle t\rangle},
\label{eq03}
\ee
where $\langle t\rangle= \mathrm{Tr} \left[\mathcal L^{-1}_0
\rho(0)\right]$ is the the average trapping time for the initial
density matrix $\rho(0)$, and $\mathcal L_0 =\mathcal L_{\mathrm{sys}}+
\mathcal L_{\mathrm{trap}}+\mathcal L_{\mathrm{dissip}}$ is the new
Liouville superoperator in the absence of decay.

\begin{figure}[htp]
  \includegraphics[width=1\columnwidth]{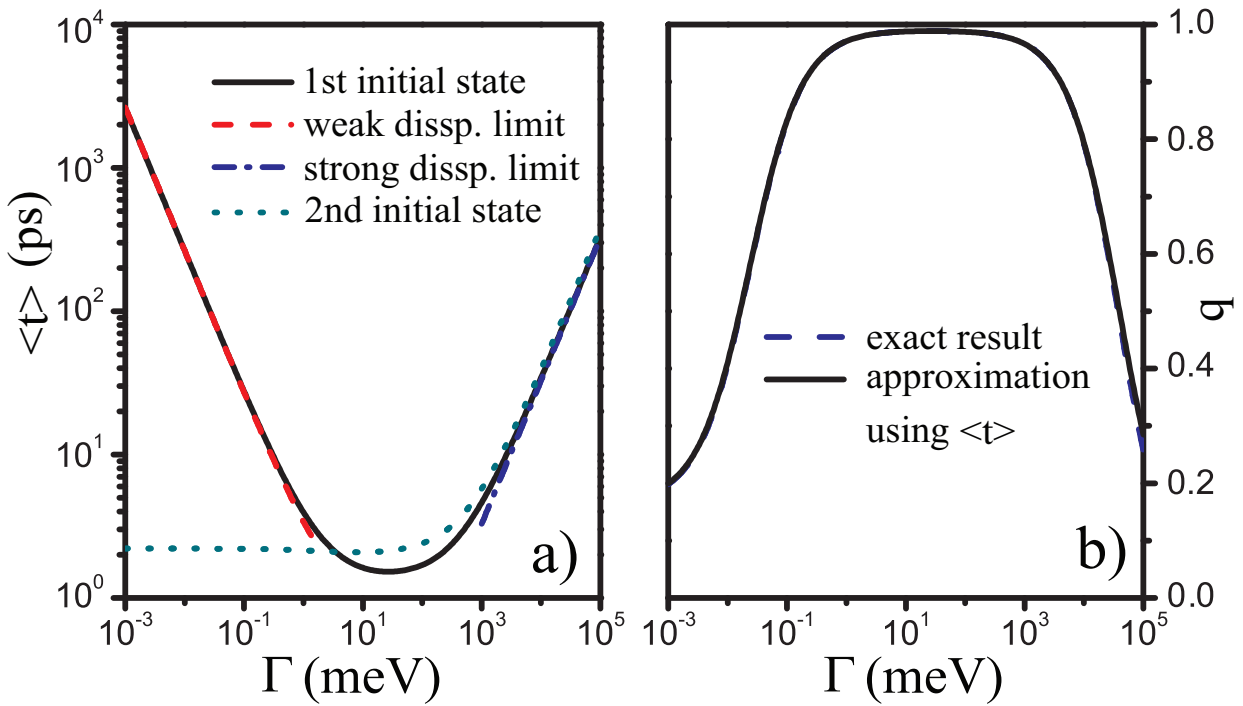}
  \caption{EET in the dendrimer depicted in Fig.~\ref{fig00}a. a) The trapping time $\langle t\rangle$ vs. the pure dephasing rate $\Ga$.
  For the incoherent initial condition defined at six outer 2nd-generation sites (i.e, the non-trapping state), the solid line is a full calculation.
  The dotted-dashed line is the large-$\Ga$ result from Eq.~(\ref{eq04a}).
  The dashed line is the small-$\Ga$ result from Eq.~(\ref{eq11}). As a comparison, the dotted line is the result
  for the coherent initial condition defined at three 1st-generation sites (i.e, the trapping state).
  b) The transfer efficiency $q$ vs. $\Ga$ for the 1st initial condition. The dashed line
  is the exact result as calculated from Eq.~(\ref{eq02}) while the solid line is obtained from the approximation in Eq.~(\ref{eq03})
  and modification for $\Ga\lesssim 1$.
  %with a modification for $\Ga\lesssim 1$ (see text).
  Here parameters are $J=20$ meV, $k_t=5$ meV, and $k_d=5$ $\mu$eV.
  }
  \label{fig02}
\end{figure}

The optimization of energy transfer efficiency is thus simplified to
the minimization of $\langle t\rangle$.  It has been found that
for many EET systems, $\langle t \rangle$
decreases with increasing $\Ga$ for weak dissipation and diverges
with $\Ga$ for strong dissipation, where $\Ga$  represents
a characteristic dissipation strength in $\mathcal L_{\mathrm{dissp}}$, %$\mathcal L_{\mathrm{dissp}}\propto \Ga$,
e.g., the dephasing rate.~\cite{gaab2004:JChemPhys,vlaming2007:JCP,caruso2009:JChemPhys,robentrost2009:NewJPhys,cao2009:JPCA,wu2010:NewJPhys}. %olaya-castro2008:PhysRevB,
A representative curve of this dependence is plotted in
Fig.~\ref{fig02}a, where the trapping time $\langle t\rangle$ is minimal
at an optimal noise level $\Ga_{\mathrm{opt}}$. To understand this
generic behavior, we investigate the asymptotic scaling of the trapping
time in the strong and weak dissipation limits.

{\it Asymptotics of the trapping time.} --- In the strong
damping limit ($\Ga \rightarrow \infty$), quantum coherence is
quickly destroyed by noise, and energy is transferred from donor
sites to the trap  through a series of incoherent hops.
The hopping rate $k_{\mathrm{hop}}$ can be estimated from Fermi's golden rule,
$k_{\mathrm{hop}}\sim |J|^2/\Ga$,
where $J$ is the exciton coupling strength.
This relation is consistent with the F\"{o}rster
theory %~\cite{forster1948,forster}
where $\Gamma$ is %understood as
proportional to the spectral line-width. The average trapping time
diverges linearly with $\Gamma$, giving \be \langle t\rangle \sim
k^{-1}_{\mathrm{hop}}\sim  \Gamma/|J|^2. \label{eq04a} \ee In a
recent paper, we mapped exciton dynamics to network kinetics using a
systematic expansion and recovered hopping kinetics as the leading
order term~\cite{cao2009:JPCA} in the strong damping limit.  %in Eq.~(\ref{eq04a}).
% where the hopping rate is inversely proportional to the friction $\Ga$.
The exact $\Ga$-dependence of the hopping rate $k_{\mathrm{hop}}$ is
usually more complicated for a quantum non-Markovian description of
dissipation. Nevertheless, the linear $\Ga$-dependence of the
trapping time in Eq.~(\ref{eq04a}) can serve as a simple scaling
relation for relative large values of damping\cite{note0}.

In the opposite weak damping limit ($\Ga\rightarrow 0$), energy transfer
can be enhanced by dynamic noise such that $\langle t\rangle$
decreases with increasing $\Ga$.
Here the starting point for studying EET
dynamics is the delocalized exciton basis, i.e., eigenstates of
$H_\mathrm S$, and energy is transferred
coherently to the trap state through delocalized excitons.
However, not every exciton state is capable of
efficient energy transfer; a set of excitons, orthogonal to the
trapping operator, define the trapping-free  subspace  $\Phi_\perp$ (i.e., orthogonal subspace),
    \be
\mathcal L_{\mathrm{trap}} \rho_{\perp} \cong 0,
\label{eq04z}
    \ee
where the index $\perp$ denotes the trapping-free subspace such that
$\rho_\perp \in \Phi_\perp$.
Without the coupling to the trap,
Eq.~(\ref{eq04z}) indicates that the population in $\Phi_{\perp}$
is conserved and therefore the coherent trapping time diverges,
$<t>|_{\Gamma=0}=\infty$. For a single trap site,
the procedure of identifying $\Phi_\perp$ has been provided
in Ref.~\cite{caruso2009:JChemPhys}.
We note that  Eq.~(\ref{eq04z}) is
analogous to  the definition of decoherence-free subspace in quantum
information, where the de-coupling from  the
system-bath interaction~\cite{whaley1998:PRL} is defined as the de-coupling
from the system-trap overlap.

The system-bath coupling induces interactions between the
trapping-free and other exciton states,  thus leading to population
depletion from the trapping-free subspace. Since the population of
the unperturbed orthogonal subspace will not deplete, dissipation of
the orthogonal subspace $\Phi_\perp$ dominates the population
transfer time to the trap, giving $\left({\mathcal
L^{-1}_0}\right)_{\boldsymbol{\perp}} \approx (\mathcal
L_{\mathrm{dissip}; \boldsymbol{\perp}})^{-1}$, where $\mathcal
L_\perp$ represents the matrix element of the Liouville
superoperator defined in $\Phi_\perp$. For nonzero population in
$\Phi_{\perp}$, the leading order of the average trapping time is
given by the survival time in the orthogonal exciton subspace,
 \be
 \langle t \rangle \approx
\mathrm{Tr} \left[ \left(l_{\mathrm{dissip; \boldsymbol {\perp}}}
\right)^{-1}\rho_{{\perp}}(0)\right] / \Ga,
\label{eq11}
\ee
where the linear form $\mathcal L_{\mathrm{dissip}} \approx \Ga
l_{\mathrm{dissip}}$ is valid in the limit of $\Ga\rightarrow 0$,
and $l_{\mathrm{dissp}}$ is the reduced Liouville superoperator, independent of $\Ga$.

{\it Generality.} --- The two scaling relations in the asymptotic
regimes are based on general physical arguments and thus independent
of specific details such as the system-bath coupling, bath spectral
density, truncation method, and approximations in the quantum master
equation. Combining Equations~(\ref{eq04a}) and~(\ref{eq11}), we
obtain an estimation of the optimal condition, \be
\Ga_{\mathrm{opt}} \sim \left\{\mathrm{Tr} \left[
\left(l_{\mathrm{dissip; \boldsymbol {\perp}}}
\right)^{-1}\rho_{{\perp}}(0)\right]\right\}^{1/2} |J|,
\label{eq11b} \ee which depends on the system Hamiltonian as well as
the initial condition. In fact, the optimal efficiency or minimal
trapping time is analogous to the Kramers turnover in reaction rate
theory, where the two scaling regimes correspond to energy diffusion
and spatial diffusion, respectively~\cite{berne,hanggi}. The
change of the reaction coordinate from energy to spatial position in
classical rate theory corresponds to the change of basis set from
excitons to local sites in energy transfer theory. However, the
analogy to classical rate theory is limited to the orthogonal
subspace and does not apply to the non-orthogonal subspace, where
coherent energy transfer to the trap state is the dominant pathway
in the weak damping limit and will not display noise-induced
enhancement as illustrated in Fig.~\ref{fig02}A.

The orthogonal condition in Eq.~(\ref{eq04z}) for $\Phi_{\perp}$ can
be realized in various systems.
(i) In an inhomogeneous system with large energy mismatches (see
Fig.~1a), the orthogonality can arise from a vanishingly small overlap
coefficient between  the donor and acceptor.
This situation can be well described by Fermi's golden rule %the F\"{o}rster theory
and qualitatively explains the optimal efficiency observed in
FMO~\cite{wu2010:NewJPhys,wu2012:Flux}. (ii) In a spatially extended
system with a large number of sites, the orthogonality can arise
because the average overlap coefficient of a local site with the
trap decreases with the system size, e.g., a long, homogeneous chain
system (see Fig. 1b). (iii) In a system with intrinsic symmetry,
each exciton state resulting from the diagonalization of
$H_{\mathrm{sys}}$ is associated with a specific symmetry. A subset
of these eigen-states are incompatible with the symmetry of the
trapping Liouville operator $\mathcal L_{\mathrm{trap}}$, thus
leading to orthogonality. The fully-connected
network~\cite{caruso2009:JChemPhys} and the dendrimer in
Fig.~\ref{fig00}c are examples of such topological symmetry. The
orthogonality due to symmetry in case (iii) is rigorous, whereas the
orthogonality in cases (i) and (ii) requires either a large energy
mismatch (i.e. detuning) or a large system size. In cases (i) and
(ii), the average trapping time may not diverge at $\Gamma=0$ but
can still lead to an optimal efficiency because of noise-enhancement
under weak non-orthogonality.

{\it An example} --- To verify the above analysis, we consider a
two-generation three-fold dendrimer depicted in
Fig.~\ref{fig00}c~\cite{supritz2006:JLumin}. A tight-bonding model
is used for the system Hamiltonian, giving $H_{\mathrm S; ij} =
\varepsilon_i \de_{ij} |i\rangle\langle i| + J_{ij}(1-\de_{ij})
|i\rangle\langle j|$, where $|i\rangle$ represents a localized
excited state at site $i$. All the site energies $\varepsilon_i$ are
assumed to be identical and the site-site interaction $J_{ij}=20$
meV is defined between two connected sites. The center site is the
trap where the trapping process occurs with rate $k_t=5$ meV. The
decay process is characterized by a homogeneous rate $k_d= 5$
$\mu$eV. For simplicity, we approximate dissipation by the
Haken-Strobl-Reineker (HSR) model~\cite{Haken1972,haken1973:ZPhys}
and consider homogeneous pure dephasing, $\mathcal
L_{\mathrm{dissip}; ij} = (1-\de_{i,j})\Gamma$, where $\Ga$ is
defined as the pure dephasing rate for each coherence element
$\rho_{ij}$. For this dendrimer system, the
symmetric Hamiltonian has an orthogonal %exciton
subspace consisting of seven exciton states. We now compare two
different cases of initial preparation. In the first case, an
incoherent population is evenly distributed at six outer
2nd-generation sites, leading to $\rho_{\perp}(0)\neq 0$. The
average trapping time is plotted as a function of the pure dephasing
rate in Fig.~\ref{fig02}a. The divergence of $\langle t\rangle$ in
the strong and weak dephasing limits is in an excellent agreement
with the asymptotic behaviors predicted in Eqs.~(\ref{eq04a})
and~(\ref{eq11}), respectively. Figure~\ref{fig02}b shows that our
approximate equation in Eq.~(\ref{eq03}) provides a quantitatively
accurate description for the EET efficiency. Here a slight
modification is applied in Eq.~(\ref{eq03}) to correct the
zero-dissipation transfer efficiency~\cite{note1}. In the second
case, a coherent state is evenly distributed at three 1st-generation
sites, leading to $\rho_{\perp}(0)= 0$. The small-$\Ga$ divergence
of $\langle t\rangle$ in Eq.~(\ref{eq11}) disappears since no
initial population exists in the orthogonal subspace. Above a
threshold at the intermediate dephasing rate, $\langle t\rangle$
changes from a plateau to the same linearly increasing function of
$\Ga$. This calculation confirms that dynamic noise can enhance the
EET if the exciton subspace $\Phi_{\perp}$ is orthogonal to the
trapping process, and demonstrates the relevance of the initial
condition for the energy transfer efficiency and thus the
possibility of quantum control of energy transfer efficiency.
%Our latest calculation in FMO using the hierarchic method has further
%confirmed the two asymptotic scalings for the relation
%between the trapping time and the reorganization energy in the Debye spectral density.

\begin{figure}[htp]
   \includegraphics[width=0.95\columnwidth]{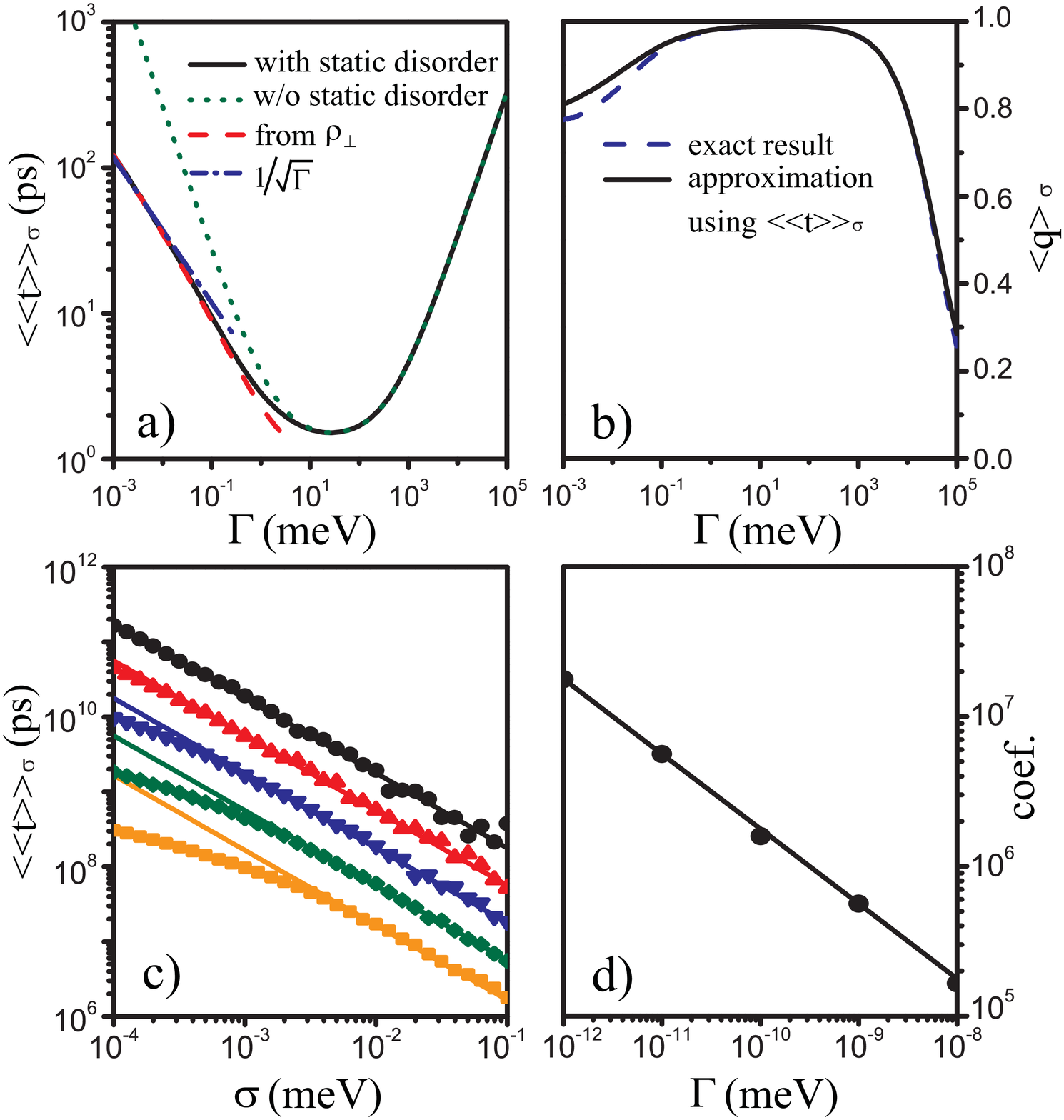}
  \caption{The ensemble-averaged EET for the same dendrimer as in Fig.~\ref{fig02} with a static disorder $\sigma=4$ meV.
  a) The solid line is the simulation result of $\langle\langle t\rangle\rangle_\sigma$,
  referenced by the dotted line without $\sigma$ from Fig.~\ref{fig02}a.
  The dashed line is the dominant contribution from the orthogonal subspace, and the dashed-dotted line
  is the fitting result of $1/\sqrt{\Ga}$. b) $\langle q\rangle_\sigma$ vs. $\Ga$: the solid line is the ensemble
  average of Eq.~(\ref{eq02}) whereas the dashed line is from Eq.~(\ref{eq03}) using $\langle \langle t\rangle\rangle_\sigma$
  and modification for $\Ga\lesssim 1$. c) The weak-dephasing $\sigma$-dependence of $\langle\langle t\rangle\rangle_\sigma$,
  with $\Gamma = 10^{-12}$,  $10^{-11}$, $10^{-10}$, $10^{-9}$, and $10^{-8}$ meV (from top to bottom).
  Symbols denote simulation results, whereas the solid lines are fitted
  with $\langle\langle t\rangle\rangle_\sigma = c^\pr /\sigma$. d) The fitting coefficient $c^\pr$ (circles)
  is plotted as a function of $\Ga$ and can be further fitted by $c^\pr=c/\sqrt{\Ga}$ (the solid line).
  }
  \label{fig03}
\end{figure}

{\it Static disorder.} --- We introduce an energy disorder $\de
\varepsilon_i$ at each site $i$, that follows the Gaussian
distribution, $P(\de\varepsilon_i) = \exp[-(\de\varepsilon_i^2/(2
\sigma^2_i))]/\sqrt{2\pi}\sigma_i$, where $\sigma_i$ is the variance
of energy disorder. The resulting ensemble average is given by
$\langle x \rangle_{\sigma} = \Pi_{i}\int  x(\de\varepsilon_i)
P(\de\varepsilon_i) d\de\varepsilon_i$ with $x=\langle t\rangle$ or
$q$. Using the above dendrimer model with the first initial
condition, we present the results of $\langle\langle
t\rangle\rangle_\sigma$ and $\langle q\rangle_\sigma$ obtained from
a Monte Carlo simulation of $10^5$ samples. As shown in
Fig.~\ref{fig03}a, static disorder is irrelevant in the strong
dephasing limit ($\Ga\gg\sigma$). %, where the effect of static disorder $\sigma$ is perturbative.
A more interesting $\sigma$-dependence occurs in the weak dephasing
limit ($\Ga\ll \sigma$), where the trapping time is dominated by the
survival time in the trapping-free %orthogonal
subspace. Static disorder can
destroy the orthogonality in Eq.~(\ref{eq04z}) and induce a large
reduction of the trapping time. However, if the random energy
disorder falls below the amplitude of stochastic noise, the weak
orthogonality is preserved and the trapping time diverges as $
\langle t \rangle \propto 1/\Gamma$. Since the amplitude of the
stochastic noise is proportional to $\sqrt{\Gamma}$, the range of
relevant random energy that contributes to the divergence is defined by
$\sigma^\pr\sim \sqrt{\Ga}$. Thus, we obtain the new asymptotic
relation in the weak dephasing limit by an integral over these small
disorders, giving
    \be
\langle\langle t \rangle\rangle_\sigma \sim \int_{-\sigma^\pr}^{\sigma^\pr}d \de\varepsilon
 P(\de\varepsilon)~ \langle t\rangle|_{|\de\varepsilon|<\sigma^\pr}
\sim \frac{c}{\sigma\sqrt{\Ga}}, \label{eq11a} \ee where
$\de\varepsilon$ describes the effective in-phase energy fluctuation
over all the sites, $P(\de\varepsilon)\sim 1/\sigma$ is the
probability distribution, and $\langle
t\rangle|_{|\de\varepsilon|<\sigma^\pr}\approx\langle
t\rangle|_{\de\varepsilon=0} \sim 1/\Ga$ is approximately uniform
within this regime. The pre-factor $c$  depends on the initial
condition, the system Hamiltonian, and the trapping rate. We confirm
the asymptotic relation, $\langle\langle t\rangle\rangle_\sigma
\propto 1/\sqrt{\Gamma}$, by numerical fitting in Fig.~\ref{fig03}a.
Further, we calculate the weak-dephasing $\langle\langle
t\rangle\rangle_\sigma$ as a function of $\sigma$ for different
values of $\Gamma$ in Figs.~\ref{fig03}c-d and rigorously estabalish
the scaling relation predicted by Eq.~(\ref{eq11a}), i.e.,
$\langle\langle t\rangle\rangle_\sigma \propto 1/\sigma$.
 Our calculations suggest that nature can use both static and dynamic disorders
cooperatively to achieve efficient and robust energy transfer.
Indeed, as shown in Fig.~\ref{fig03}b, the zero dephasing efficiency
$\langle q|_{\Ga=0}\rangle_\sigma$ is drastically enhanced from 0.2
to 0.8 with $\sigma=J/5$.
% and the asymptotic scaling of efficiency changes from
%$\delta\langle q \rangle \propto \Gamma$ without disorder to $\delta\langle q
%\rangle \propto \sqrt{\Gamma}$ in the weak disorder regime.

{\it Conclusion.} --- In this Letter, we demonstrate that the
generic mechanism of  noise enhanced EET is to assist energy flow
out of the orthogonal exciton subspace, and the competition between
noise-enhanced EET in the weak dissipation regime and noise-induced
localization in the strong dissipation regime leads to an optimal
efficiency. We determine the scaling relations of the average
trapping time in these two regimes and use the asymptotic relations
to qualitatively predict the optimal noise. The presence of static
disorder reduces the exponent of divergence in the weak-dissipation
limit and thus makes the EET process more robust against noise. Our
analysis is not limited to EET but also applies in general to
electron transfer, %~\cite{marcus},
fluorescence emission, heat conduction and other open quantum
processes. Sophisticated numerical methods, such as the hierarchy
equation approach, will be used to quantify the general optimal
condition in the EET process.~\cite{note0}

This work was supported by grants from the National Science
Foundation (Grant CHE-1112825) and DARPA (Grant N66001-10-1-4063).
JC is partially supported by the Center for Excitonics funded by the US Department of Energy
%an Energy Frontier Research Center funded by the
%US Department of Energy, Office of Basic Energy Sciences
(Grant DE-SC0001088). JW acknowledges
partial support from the Fundamental Research Funds for the
Central Universities in China (Grant 2011QNA3005)
and the National Science Foundation of China (Grant 21173185).


\begin{thebibliography}{15}
\bibitem{note0}
{ We have obtained the detailed $\Ga$-dependence of the trapping time in
the two-state system and the FMO system using the hierarchy equation
approach. Our results agree with the two scaling relations within the physically reasonable range of $\Ga$.}
\bibitem{note1}
For small dephasing rates, $\langle t\rangle$ is
replaced by $1/(\langle t\rangle^{-1}+t^{-1}_0)$ in Eq.~(\ref{eq03}) where
the constant $t_0=(q|_{\Ga=0}^{-1}-1)/k_d$ corrects the zero-dissipation transfer efficiency.
\end{thebibliography}
\end{document}